\documentclass[preprint]{aastex}

\shorttitle{Source Redshift of B0218+357}

\newcommand{\blazar}{B$0218+357$}
\newcommand{\kms}{km~s$^{-1}$}

\newcommand{\etal}{{\it et al.\/}}

\begin{document}

\title{The Redshift of the Lensed Object in the Einstein Ring B0218+357
\altaffilmark{1}}

\author{Judith G. Cohen \altaffilmark{2}, Charles R. Lawrence\altaffilmark{3}
and Roger D. Blandford\altaffilmark{4} }

\altaffiltext{1}{Based on observations obtained at the
W.M. Keck Observatory, which is operated jointly by the California
Institute of Technology, the University of California, and the
National Aeronautics and Space Administration.}

\altaffiltext{2}{Palomar Observatory, Mail Stop 105-24,
California Institute of Technology.}

\altaffiltext{3}{California Institute of Technology, 
Jet Propulsion Laboratory, M/S 
169-327, 4800 Oak Grove Drive, Pasadena, CA 91109}

\altaffiltext{4}{California Institute of Technology, 130-33, Pasadena,
California, 91125}

\begin{abstract}

We present a secure redshift of $z=0.944\pm0.002$ for the 
lensed object in the Einstein ring gravitational
lens \blazar\ based on five broad emission lines, in
good agreement with our preliminary value announced several
years ago based solely on the detection of a single emission line.

\end{abstract}


\section{Introduction}

\blazar\ is a strong flat-spectrum radio source 
which was discovered to be gravitationally
lensed by \cite{pat93}. The object consists of
two point sources which
are radio loud and variable, with similar flat spectra in the radio
regime.  In addition to the point sources, there is an Einstein
ring of 0.33 arcsec diameter. This object is the
smallest known Einstein radio ring (Patnaik \etal\ 1993).
\cite{big99} (see also Cohen \etal\ 2000) have measured
a time delay for \blazar\ of 10.5$\pm0.4$ days.
HST optical images by \cite{kee98} and by \cite{leh00} and NICMOS images
by \cite{jac00} reveal
the lensing galaxy clearly.

\blazar\ is an extremely important object for studies of
gravitational lensing.  The Einstein ring provides a strong
constraint on the mass distribution in the lens.
The simple source structure and small angular size of \blazar,
the constraints from sub-structure in the
images which are well mapped with VLBI 
\citep*[see][]{biggs2002}, and
the apparent absence of significant
external shear make this object easier than most to model.
As the error on the time delay is 
currently 
estimated to be 3 percent
(1$\sigma$), (Biggs, private communication), \blazar\ is a prime target 
for determining the Hubble constant and so it is vital to obtain secure 
and accurate lens and source redshifts.

In order to fulfill the promise of \blazar\ for these purposes,
both the source and the lens redshift are required.  \cite{bro93},
and independently \cite{sti93},
established the redshift of one of these, which they assumed
was the lens, as $z=0.6847$, detecting several emission and absorption
features.   Subsequently
absorption arising in the lensing galaxy
has been detected at 21 cm \citep{car93},
in CO \citep{wik95}, and in formaldehyde \citep{men96}
against the background source.

However, not surprisingly, the source redshift for \blazar\ proved
more elusive, since the source is a blazar \citep{odea92}.  We therefore
started an effort to find the source redshift in 1994 using
the Low Resolution Imaging Spectrograph \citep{oke95}
at the Keck Observatory.
We were able to detect one definite weak emission feature fairly
rapidly, but could not find a secure second line.  Assuming the detected
line at 5460\AA\ is the 2800 \AA\ Mg II line, the redshift of the 
lensed object in \blazar\ is then $z \sim0.95$, further supporting
\cite{bro93} and \cite{odea92}, who offer $z=0.94$ as a
``tantalizing possibility, rather than
a firm claim''.
This result was announced at two conferences in 1996
\citep{coh96,law96}.
It has taken longer than expected,
but we have finally succeeded in detecting
with confidence multiple emission lines from the source.
We present in this brief paper the secure
redshift of the lensed object in \blazar\ based on the
detection of five broad emission lines, $z=0.944\pm0.002$.

\section{Observations}

\begin{figure}
\epsscale{1.0}
\plotone{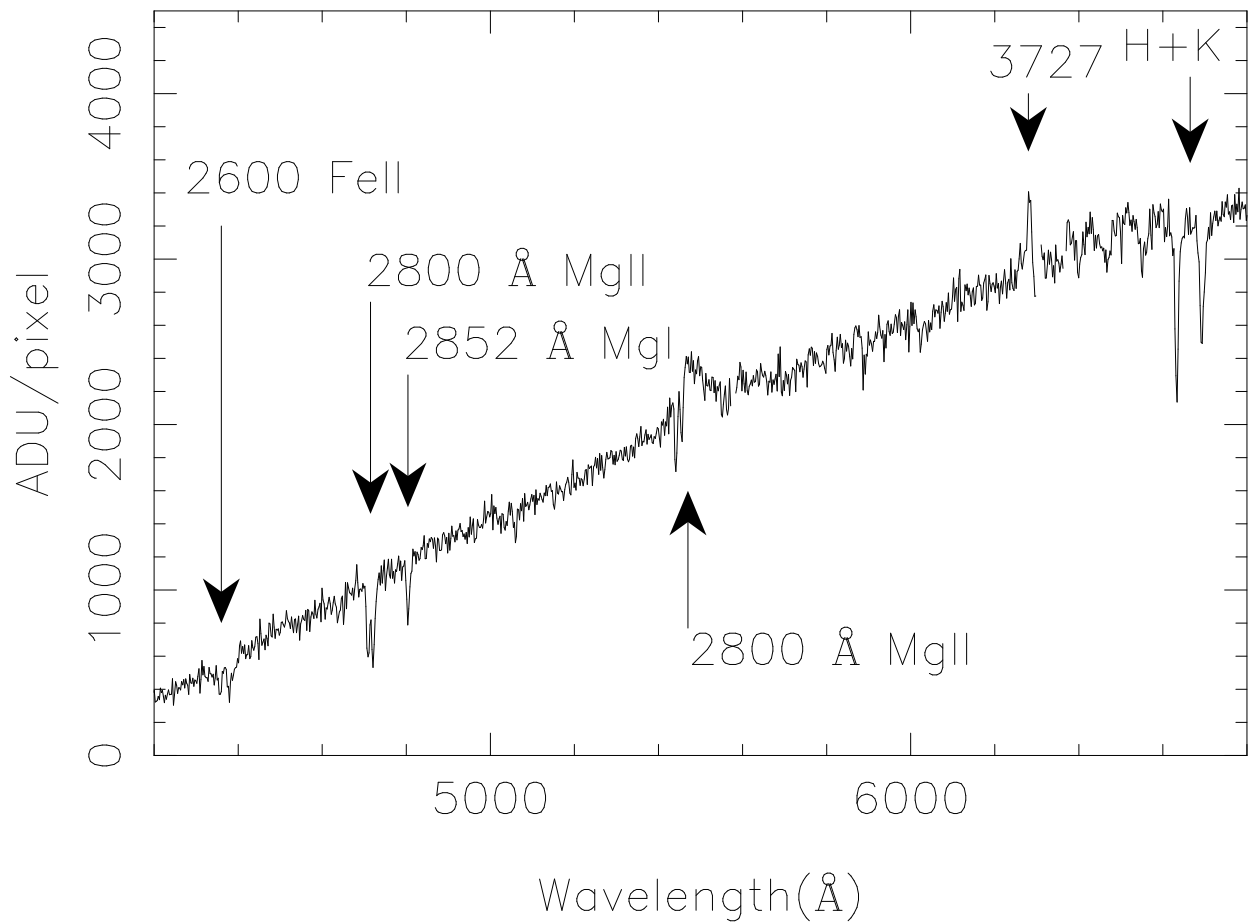}
\caption[]{The blue spectrum of \blazar.  This is the sum
of four 1000 second exposures taken with LRIS in Feb. 1997.  The lines from
the lens are labeled above the spectrum, while that from the lensed 
object (the 2800 \AA\ Mg II line)
is labeled below the spectrum. Regions where the night sky
emission lines are strong have been omitted.
\label{fig_blue}}
\end{figure}

In August 1994, we obtained several exposures with the LRIS of \blazar,
with a total exposure time of 6200 sec.  The 300 g/mm grating
was used and the spectra covered the wavelength
region 3000 to 7600 \AA\ (with no useful signal below 4000\AA).
Figure~\ref{fig_blue} shows the spectrum of
\blazar\ over the central part of
this wavelength regime (from spectra with the same
instrumental configuration taken in 1997).
The steep spectral slope is apparent in this figure. The
optical spectrum is fairly red;
$F_{\nu} \propto \nu^{\alpha}$, with $\alpha \sim -3.5$ \citep{odea92}.
Numerous absorption lines and a few emission lines from the lensing
galaxy are clearly seen at the expected redshift, $z=0.68$
\citep{bro93}.  In addition, we found a single
broad emission line at $\sim$5470\AA\ which has
absorption components on its blue wing.
We could not find a second emission line within the
spectral range covered, and hence the most
likely line identification for the detected broad line
is Mg~II at 2800\AA, implying $z$ for
the lensed object is $\sim$0.95.

If this value is correct, then the next strong emission line to
the blue should be  CIII] 1909~\AA\  expected to
be at $\sim$3720~\AA, while H$\beta$ should be at about 9500~\AA.
Since the blue side of LRIS was not installed until 2001,
looking for the CIII] line was impossible.  Searching for the
Balmer lines was also not trivial because these are likely to be weak broad
features, and the night sky emission and absorption will create difficulties,
but this was judged the only feasible route.

We took spectra of \blazar\ in the red in 1995,
1996, 1997 and 2000 to hunt for these lines
in order to confirm our preliminary redshift.  A number of hot stars
(white dwarfs or rapidly rotating O stars) selected
from \cite{oke90} with broad
spectral features (and no strong ones within the red region
we use here) were observed at a similar range of airmass each night
to enable removal
of terrestrial atmospheric absorption bands.
The object was slightly dithered by several arcsec
along the slit between exposures.  The 600 g/mm grating blazed at 7500~\AA\
was used, with spectral coverage from 7530 to 10,100~\AA.
A sophisticated reduction and analysis
scheme was developed, using the Laplacian cosmic ray removal
algorithm of \cite{dok01}, followed by standard Figaro
scripts \citep{sho93} for removing distortions in
two dimensions, as well as subtraction of the resulting frames.
Because both the emission and absorption night sky spectrum are
temporally variable (as well as dependent
on airmass), this did not result in perfect removal of the
night sky emission lines, but it did reduce their intensity by
a large factor.
This was followed by a removal of the residual night sky emission
and, as well as possible, the night sky absorption.

\begin{figure}
\epsscale{1.0}
\plotone{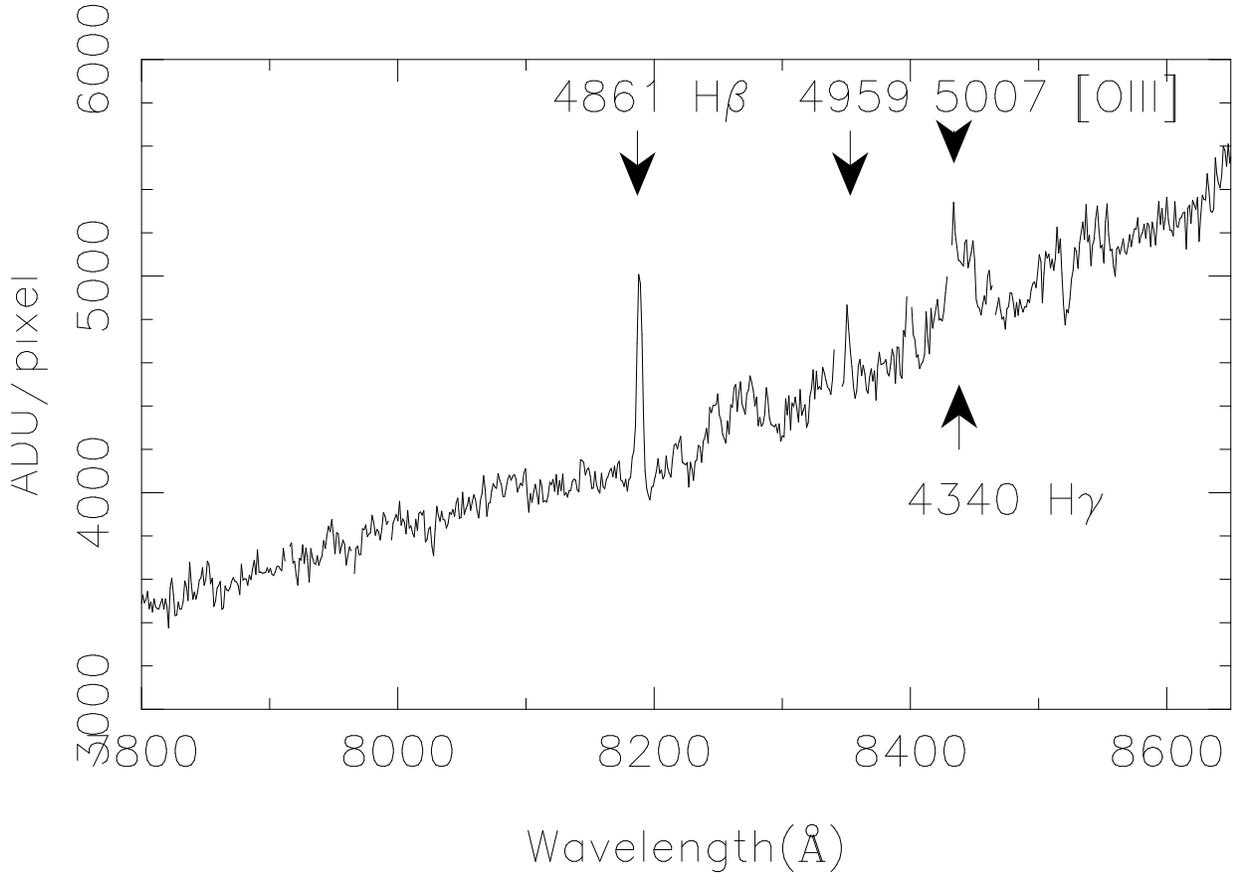}
\caption[]{The spectrum of \blazar\ from 7800 to 8650 \AA.
This is the sum of
exposures taken with LRIS in 1995, 1996 and 1997 with a total
exposure time of 16,500 sec. The
spectrum has been divided by an exposure of the hot star
G191-B2B to remove atmospheric absorption features.
As in Fig.\ref{fig_blue},
the lines from
the lens are labeled above the spectrum, while the H$\gamma$ emission 
from the lensed object is labeled below the spectrum. 
Regions where the night sky
emission lines are strong have been omitted.
\label{fig_red85}}
\end{figure}

Figure~\ref{fig_red85} shows the region from 7800 to 8650~\AA.
The sum of the 1995, 1996 and 1997 red spectra of \blazar\
with a total exposure time of
16,500 sec is used.   (Several additional spectra of this object taken
during that period through thin clouds with lower signal levels
were not included.)  The summed spectra have been divided by a
spectrum of the DA0 white dwarf G191B2B, whose signal level
is much higher than that of the spectrum of \blazar.  The
spectrum of G191B2B was normalized to the
region near 8800~\AA.  Prior to this division, the summed, sky-subtracted
spectrum of \blazar\  had $\sim$8500 ADU/pixel (with 2 $e^{-}$/ADU)
in the continuum at 8100\AA, and $\sim$2000 ADU/pixel in the continuum at
9800\AA.
Regions within the strongest night
sky lines are omitted in the figure; at such points the
curve is discontinuous.  This spectral region shows
a strong narrow emission line (H$\beta$ from the lensing galaxy)
plus a weaker broad emission line  which we interpret as H$\gamma$
in the background source.  The blue wing of the latter 
is interrupted by the strong
8430\AA\ night sky emission line.    The 4959, 5007~\AA\ [OIII]
doublet  from the lens is  present. The  bluer line of the doublet
is slightly to the red of the
strong night sky
emission line at 8343\AA\ (where the curve is discontinuous), while
the redder line is
mixed in as a sharp peak within the broader H$\gamma$ from the background
source.
Small residuals of features from the night sky are also apparent.

\begin{figure}
\epsscale{1.0}
\plotone{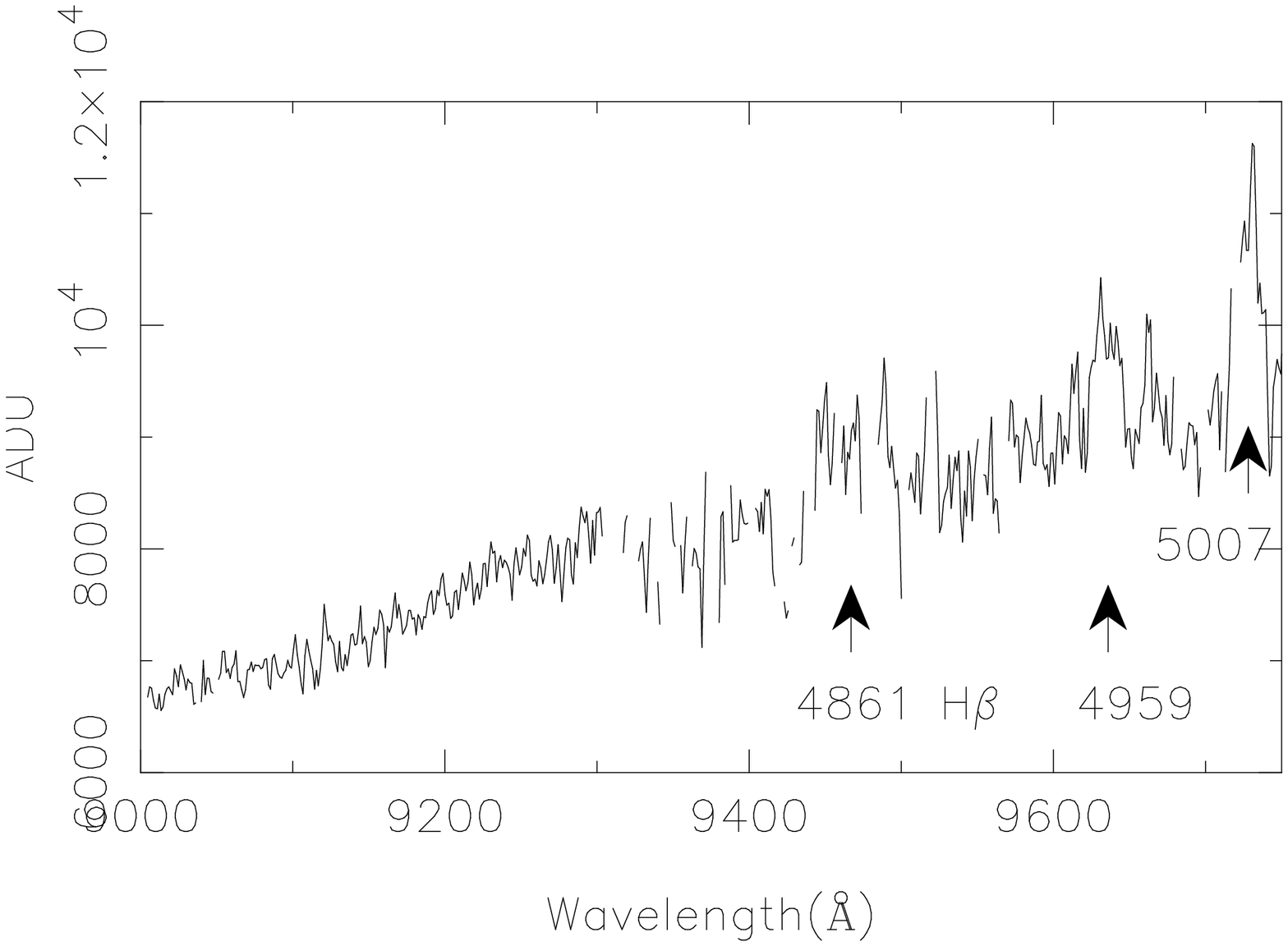}
\caption[]{The spectrum of \blazar\ from 9000 to 9750~\AA.
This is a redder section of the same spectrum as shown in Fig.~\ref{fig_red85}.
In addition, here the atmospheric absorption bands, which are very
strong from 9300 to 9500~\AA, were removed by
a scaling of the spectra of bright hot stars.
Regions where the night sky
emission lines are strong or where the hot star spectrum is
changing rapidly ($>8$\%/pixel) have been omitted.  The same convention
is followed of
labelling lines from the lens galaxy above the spectrum and lines
from the lensed object below the spectrum.
\label{fig_red95}}
\end{figure}

The region of H$\beta$ in the source is shown in
Figure~\ref{fig_red95}.  Even though this is a total of
16,500 sec exposure of \blazar, the signal is weak
due to the decrease in LRIS instrumental efficiency
at such red wavelengths.  As before,
regions where the night sky
emission lines are strong have been omitted.
The
spectrum has been divided by an exposure of the hot star
G191-B2B to remove atmospheric absorption features
normalized as described  above.
In this spectral region from 9300 to 9500~\AA,
the night sky absorption is large and rapidly varying.
Regions where the hot star spectrum changed by more than 8\%/1.2\AA\ 
(1 pixel) are also omitted.  Furthermore, the absorption in the hot
star is slightly scaled to better match the absorption
in \blazar.  In addition to H$\beta$,
the somewhat narrower, but
still broad (compared to the lines arising from the lens)
lines of the 4959, 5007~\AA\ [OIII] doublet  from the source are apparent.
Strong, narrow Na D absorption lines from the lens are seen easily 
in the full spectrum, just redward of the portion shown 
in Figure~\ref{fig_red95}.

Table~\ref{tab_lines} summarizes the detected lines from both
the source and the lens in \blazar.  
Adopting the mean redshift from the four broad emission lines,
(excluding MgII 2800, which is distorted by absorption features), we
find $z=0.944\pm0.002$ for the redshift of the lensed object in \blazar.

\subsection{Internal Motions in the Lens}

Since the red spectra have higher spectral resolution,
we have used the narrow H$\beta$ emission line in the lens,
detected at 8189~\AA, to set an upper limit to the
velocity dispersion and/or rotation in the lens galaxy.
The instrumental resolution (projection of a 1.0 arcsec wide slit)
of LRIS
corresponds at that wavelength to 204 \kms.  We determine an
upper limit to the contribution to the FWHM line width from internal
motions in the lens galaxy of $\le 100$ \kms.
\cite{pisano01} have studied the relationship
between the width of H$\beta$ emission and of 21 cm emission
in a sample of galaxies, and find that this ratio has
a mean of about 0.7, presumably because the ionized gas which
gives rise to the former is more centrally concentrated within
the gravitational potential of the galaxy.

Our upper limit is consistent with
the measured FWHM of the 21 cm absorption
line of \cite{car93} of only 43 \kms.  However, the 21 cm
absorption presumably 
probes only a small pencil beam through the lensing galaxy, and does not
sample its full rotation curve.

The spectral features we detect are 
consistent with the lens being a spiral galaxy.

\section{Comment}

Given the very heavy use the community has made
of our preliminary redshift for \blazar,
published in 1996 \citep{coh96,law96}, it is indeed fortunate
that the definitive redshift of $z=0.944$ presented here confirms our
earlier (educated) guess.

\acknowledgements
The entire Keck/LRIS user community owes a huge debt to
Jerry Nelson, Gerry Smith, Bev Oke, and many other
people who have worked to make the Keck Telescope and LRIS
a reality and to operate and maintain the Keck Observatory.
We are grateful to the W. M.  Keck Foundation for the vision to fund
the construction of the W. M. Keck Observatory.
The authors wish to extend special thanks to those of Hawaiian ancestry
on whose sacred mountain we are privileged to be guests.
Without their generous hospitality, none of the observations presented
herein would have been possible.  The extragalactic
work of JGC is not supported by any federal agency.
RDB acknowledges support by the National Science Foundation under grant
AST-9900866.

\clearpage


\clearpage

%
%
\clearpage
\begin{deluxetable}{lllcc}
\tablenum{1}
\tablewidth{0pt}
\tablecaption{Features in the Spectrum of \blazar
\label{tab_lines}}
\tablehead{\colhead{Observed $\lambda$} & \colhead{Rest $\lambda$}
& \colhead{ID} & \colhead{Line Type\tablenotemark{a}} &
\colhead{Redshift} \\
\colhead{} & \colhead{(\AA)} & \colhead{(\AA)} & \colhead{} & \colhead{}
}
\startdata
Lens: \\
4357   &    2586 & FeII  & a & 0.6848 \\
4378   &   2599 & FeII & a & 0.6845 \\
4709   &    2797 & MgII (blend) & a & 0.6836  \\
4721    &   2803 & MgII & a & 0.6843 \\
4804    &   2852 & MgI  & a & 0.6844 \\
6279   &    3727 & [OII] (blend) & e & 0.6847 \\
6625    &   3933.7 & CaII & a & 0.6842 \\
6684    &   3968.5 & CaII & a & 0.6843 \\
8188.5  &   4861.3 & H$\beta$ & e & 0.6844 \\
8352 & 4959 & [OIII] & e & 0.6842 \\
8435 & 5007 & [OIII] & e & 0.6846 \\
9922 & 5889 & Na I & a & 0.6846 \\
9933 & 5895 & Na I & a & 0.6847 \\
  ~ ~  \\
Source: \\
5466 & 2800 & Mg II & a & 0.952\tablenotemark{b} \\
8439 & 4340 &  H$\gamma$ & e & 0.944 \\
9466 & 4861 & H$\beta$ & e & 0.947 \\
9634 & 4959 & [OIII] & e &  0.943 \\
9728 & 5007 & [OIII] & e &  0.943 \\
\enddata
\tablenotetext{a}{``a'' denotes absorption features, ``e'' emission lines.}
\tablenotetext{b}{Absorption distorts the profile of the blue wing
of this line.}
\end{deluxetable}
\clearpage


\clearpage

\clearpage

\begin{thebibliography}{}

\bibitem[Biggs et al.(1999)]{big99} Biggs, A.~D., Browne, I.~W.~A.,
Helbig, P., Koopmans, L.~V.~E., Wilkinson, P.~N. \& Perley, R.~A. 1999,
\mnras, 304, 349

\bibitem[Biggs et al.(2002)]{biggs2002} Biggs, A.~D., Wucknitz, O., Porcas,
R.~W., Browne, I.~W.~A., Jackson, N.~J. \& Wilkinson, P.~N., 2002,
\mnras\ (in press) (Astro-ph/0209182)

\bibitem[Browne et al.(1993)]{bro93}
Browne, I.~W.~A.,  Patniak, A.~R., Walsh, D. \& Wilkinson, P.~N. 1993,
\mnras, 263, L32

\bibitem[Carilli, Rupen \& Yanny(1993)]{car93}
Carilli, C.~L., Rupen, M.~P. \& Yanny, B. 1993, \apj, 412, L59

\bibitem[Cohen et al.(2000)]{coh00}
Cohen, A.~S., Hewitt, J.~N., Moore, C.~B. \&
Haarsma, D.~B. 2000, \apj, 545, 578

\bibitem[Cohen(1996)]{coh96} Cohen, J.~G. 1996, in
{\it{Clusters, Lensing,
and the Future of the Universe}},
V. Trimble and A. Reisenegger (eds.), ASP Conference Series, Vol. 88,
pg. 68

\bibitem[Jackson, Xanthopoulos \& Browne(2000)]{jac00}
Jackson, N., Xanthopoulos, E. \& Browne, I.~W.A. 2000, \mnras, 311, 389

\bibitem[Keeton, Kochanek \& Falco(1998)]{kee98}
Keeton, C.~R., Kochanek, C.~S. \& Falco, E.~E. 1998, \apj, 509, 561

\bibitem[Lawrence(1996)]{law96} Lawrence, C.~R. 1996, in
{\it{Astrophysical Implications of Gravitational Lensing}},
IAU Symposium 173, Kluwer Academic Publishers, Dordrecht, pg 299

\bibitem[Lehar et al.(2000)]{leh00}
Lehar, J., Falco, E.~E., Kochanek, C.~S.,
McLeod, A., Munoz, J.~A., Impey, C.~D., Rix, H.-W.,
Keeton, C.~R. \& Peng, C.~Y. 2000, \apj, 536, 584

\bibitem[Menton \& Reid(1996)]{men96}
Menton, K.~M. \& Reid, M.~J. 1996, \apj, 465, L99

\bibitem[O'Dea et al.(1992)]{odea92}
O'Dea, C.~P., Baum, S., Stranghellini, C., Dey, A., van Breugel, W.,
Deusutua, S. \& Smith, E.~P. 1992, \aj, 104, 1320

\bibitem[Oke(1990)]{oke90} Oke, J.~B., 1990, \aj, 99, 1621

\bibitem[Oke et al.(1995)]{oke95}
Oke, J.~B.,  Cohen, J.~G., Carr, M., Cromer, J.,
Dingizian, A., Harris, F.~H., Labrecque, S., Lucinio, R., Schaal, W.,
Epps, H., \& Miller, J. 1995, \pasp, 107, 307


\bibitem[Patniak et al.(1993)]{pat93}
Patniak, A.~R., Browne, W.~A., King, L.~J., Muxlow, T.~W.~B.,
Walsh, D. \& Wilkinson, P.~N. 1993, \mnras, 261, 435

\bibitem[Pisano et al.(2001)]{pisano01}
Pisano, D.~J., Kobulnicky, H.~A., Guzman, R., Gallego, J. \&
Bershady, M.~J., 2001, \aj, 122, 1194

\bibitem[Shortrige(1993)]{sho93}
Shortridge K. 1993, in {\it{Astronomical Data Analysis Software and
Systems II}}, A.S.P. Conf. Ser., Vol 52, eds. R.J. Hannisch,
R.J.V. Brissenden \& J. Barnes, 219.

\bibitem[Stickel \& Kuhr(1993)]{sti93}
Stickel, M. \& Kuhr, H. 1993, \aaps, 101, 521

\bibitem[van Dokkum(2001)]{dok01} van Dokkum, P.~G. \pasp, 113, 1420

\bibitem[Wiklind \& Coombes(1995)]{wik95}
Wiklind, T. \& Coombes, F. 1995, A\&A, 299, 382

\end{thebibliography}
\end{document}